\documentclass[10pt]{article}

\usepackage{amsmath}
\usepackage{amssymb}
\usepackage{multirow}
\allowdisplaybreaks

\usepackage{graphicx}

\usepackage[usenames]{color} 

\usepackage{setspace} 
\usepackage{url}

\usepackage{multirow}
\usepackage{dsfont}
\usepackage{verbatim}
\usepackage{longtable}
\usepackage{esvect}
\usepackage{mathrsfs}
\usepackage{algorithm}
\usepackage{algpseudocode}
\usepackage{pifont}

\def\N{\mathcal N}
\DeclareMathOperator*{\E}{\mathbb{E}}
\usepackage{multicol}
\begin{document}

\begin{flushleft}
{\Large
\textbf{Nonparametric Reduced-Rank Regression \\for Multi-SNP, Multi-Trait Association Mapping}
}
\bigskip
\\
Ashlee Valente$^{1}$,
Geoffrey Ginsburg$^{1}$,
Barbara E Engelhardt$^{2}$
\bigskip
\\
\bf{1} Center for Applied Genomics and Precision Medicine, Duke University, Durham, NC 27708, USA
\bf{2} Department of Computer Science and Center for Statistics and Machine Learning, Princeton University, Princeton, NJ, USA
\\
\bigskip
$\ast$ Corresponding author e-mail: bee@princeton.edu
\end{flushleft}

\begin{abstract}

\noindent Genome-wide association studies have proven to be essential for understanding the genetic basis of disease. However, many complex traits---personality traits, facial features, disease subtyping---are inherently high-dimensional, impeding simple approaches to association mapping.
We developed a nonparametric Bayesian reduced rank regression model for multi-SNP, multi-trait association mapping that does not require the rank of the linear subspace to be specified. We show in simulations and real data that our model shares strength over SNPs and over correlated traits, improving statistical power to identify genetic associations with an interpretable, SNP-supervised low-dimensional linear projection of the high-dimensional phenotype. On the HapMap phase 3 gene expression QTL study data, we identify pleiotropic expression QTLs that classical univariate tests are underpowered to find and that two step approaches cannot recover.
Our Python software, BERRRI, is publicly available at GitHub: {\tt https://github.com/ashlee1031/BERRRI}
\end{abstract}

\section*{Introduction}

Genetic variation is an important driver of variability in complex  traits, contributing to phenotypic differences among individuals and across populations. To find genetic variants that regulate complex heritable phenotypes, genome-wide association studies (GWAS) have been used to identify associations between single nucleotide polymorphisms (SNPs) and quantitative traits such as height, weight, or blood pressure.

The statistical methods used for association mapping with organismal quantitative traits can also be used to identify genetic variation affecting cellular traits such as RNA expression levels or collections of co-regulated or correlated transcripts (expression QTLs or eQTLs)~\cite{brem2002genetic, zhu2008integrating, cheung2009genetics}. In this framework, the quantitative traits are the expression of RNA transcripts. eQTL analyses have been used to suggest possible biological mechanisms for trait associated SNPs (TASs) identified using GWAS, because of evidence that genetic regulation of gene expression is a primary mechanism of the genetic regulation of disease risk~\cite{nicolae2010trait,Gamazon2015}.
The availability of eQTL data for the target tissue of a complex disease provides insight into possible biological basis for disease risk and can help identify networks of genes involved in disease pathogenesis~\cite{cookson2009mapping, franke2009eqtl,he2013sherlock}.

Typically, eQTLs are classified as cis- or trans-acting, meaning the regulatory target of the QTL is allele-specific or not, respectively. 
Gene regulatory networks suggest that trans-eQTLs might be due to cis-eQTLs regulating RNA that, in turn, affect gene expression of a distal gene in a dosage-specific way~\cite{philip2014dissection}. 
A polymorphism affecting a regulatory gene in cis could thus have observable downstream effects on the expression of its targets, resulting in \emph{pleiotropic} cis- and trans-effects. Current methods identify pleiotropic eQTLs in one of three ways: i) identifying co-expressed transcripts~\cite{zhang2010bayesian, philip2014dissection}; ii) using cis-eQTLs to identify trans-eQTLs~\cite{2013arXiv1310.4792G}; or iii) projecting gene expression data down to a low dimensional space and performing association mapping in this space~\cite{Gao2014a,Gao2014}. In this paper, we perform simultaneous supervised projection of both the SNPs and the high-dimensional phenotype and association mapping using reduced rank regression.

\subsection*{Statistical challenges of eQTL analyses}

Conventional eQTL analyses identify SNPs that explain a significant fraction of the variation in gene expression levels by testing for association of expression levels of a single gene with a single SNP; in these studies, gene expression levels are treated as quantitative traits~\cite{cookson2009mapping, nica2013expression}. The univariate approach to association mapping in eQTL studies is plagued by limited statistical power to detect associations with small effects due to the large number of tests relative to the small sample sizes in these studies~\cite{stranger2011progress}. 

Cis-eQTLs have been more successfully identified than trans-eQTLs~\cite{Grundberg2012}; practically, cis-eQTLs are assumed to lie within $200$Kb of the transcription start or end site (TSS,TES) of a gene, reducing the number of tests for association by three orders of magnitude~\cite{Pickrell2010}. In contrast, interrogating the entire genome for regulatory associations for each gene---as is done for trans-eQTLs---incurs a substantial multiple hypothesis testing penalty. It is unclear if the overrepresentation of discovered cis-eQTLs is due to lower statistical power to identify trans-eQTLs, or if this overrepresentation reflects different relative numbers and effect size distributions across cis- and trans-eQTLs~\cite{wray2007evolutionary, wittkopp2008regulatory, nica2013expression,Small2011}.
However, recent studies have shown that, with sufficient sample sizes, hundreds of trans-eQTLs may be found and replicated within a tissue type~\cite{franceschini2012discovery, grundberg2012mapping}.

\subsection*{Multi-trait association mapping}

One approach to mitigate problems resulting from limited sample size is to perform association mapping across multiple traits simultaneously~\cite{sul2013, flutre2013statistical, foygel2012nonparametric, richardson2010bayesian}. The idea behind multi-trait association mapping is that the statistical tests across multiple correlated traits (such as coronary heart disease and myocardial infarction, or gene transcripts with correlated expression levels) with shared underlying genetic associations will share strength across traits to improve statistical power~\cite{stephens2013}. 
These approaches have been effective at identifying small signals present over various measured traits. For example, the Genotype-Tissue Expression (GTEx) study data includes gene expression levels and genotypes for a large number of tissue samples from hundreds of individuals~\cite{ardlie2015genotype}. It is often the case that a cis-eQTL exists in multiple tissues. In GTEx, sharing strength across many tissues identifies many more ubiquitous cis-eQTLs, or cis-eQTLs that are shared across tissues, than univariate methods alone~\cite{flutre2013statistical}. 
This approach may also be applied to correlated phenotypes~\cite{stephens2013}, medical imaging traits such as fMRI data~\cite{vounou2012sparse}, and many other high-dimensional biological and clinical traits.

Multi-SNP analyses identify a subset of local SNPs as cis-eQTLs using various types of model selection methods~\cite{Engelhardt2014, wilson2010bayesian, richardson2010bayesian,guan2011,Zhao2014}. These approaches evaluate associations with all SNPs simultaneously, often using a form of multivariate regression that includes sparsity-inducing priors on the effect sizes~\cite{tibshirani1996regression,li2010bayesian}. The many correlated SNPs (due to linkage disequilibrium) included in these models relative to sample size makes classical sparse regression techniques like Lasso non-robust~\cite{guan2011}; model averaging has been used used to create a robust multi-SNP mapping method~\cite{valdar2009mapping}.

\subsection*{Reduced Rank Regression (RRR)}

In regression models for GWAS or eQTL analyses, when the number of predictors is much greater than the number of observations (the $p \gg n$ scenario), accurately estimating the regression coefficients is a challenge~\cite{ma2014adaptive,foygel2012nonparametric, West2003}. An alternative approach to inducing sparsity in the effect sizes is to project these high-dimensional coefficients down to a lower-dimensional basis, making estimation easier because of the many fewer statistical tests~\cite{anderson1951estimating,ma2014adaptive,foygel2012nonparametric}.

Reduced rank regression (RRR)~\cite{anderson1949estimation, anderson1951estimating} does this by explicitly mapping the coefficients of a linear regression model down to a small number of latent factors. These multivariate regression models have been extended to multi-trait RRR models, which capture linear relationships between multiple response variables and multiple predictor variables on a low-dimensional linear manifold, sharing signal strength across co-regulated response variables. Inducing sparsity on that manifold further reduces the number of model parameters from standard multi-SNP, multi-trait regression models and leads to interpretable results~\cite{chen2012sparse, foygel2012nonparametric}. 

Canonical correlation analysis (CCA) also achieves this projection onto a lower-dimensional space to examine relationships between two high-dimensional data sets~\cite{Zhao2014}. However, such approaches are best suited for exploratory data analysis as they do not lead to a measure of effect size for associations in the data or a posterior probability of SNP association~\cite{Engelhardt2014}.

RRR models have previously been developed for and applied to genomic study data. For example, a sparse RRR method was developed to identify predictive signatures of disease status in fMRI images from Alzheimer's disease populations~\cite{vounou2010discovering,vounou2012sparse}. A second  RRR method was developed for genetic association studies~\cite{geweke1996bayesian, marttinen2014assessing}, and applied to data from the Northern Finland Birth Cohort to uncover two novel loci associated with lipoprotein profiles~\cite{marttinen2014assessing}.

\section*{Approach}

Here we describe our model, Bayesian extendable RRR with an Indian Buffet Process prior (BERRRI), which goes beyond previous RRR approaches in three ways to enable genome-scale association mapping. First, our model uses a nonparametric prior on the reduced rank parameters, allowing the rank to be estimated from the data. Second, BERRRI uses binary values to indicate SNP inclusion in the low-dimensional subspace, leading to interpretable posterior probabilities of association for every SNP. Third, BERRRI uses a variational Bayes approach, making posterior estimates computationally tractable for large genomic studies.

\subsection*{Bayesian Extendable RRR Model}

We propose a reduced rank regression model to quantify associations between SNPs and high-dimensional, correlated traits where the response matrix $Y\in \Re^{N \times P}$ is a set of $P$ traits for $N$ individuals. The genotype data, $X \in \Re^{N \times Q}$ for $N$ individuals and $Q$ SNPs, are encoded by $x_{n,q} \in \{0,1,2\}$ representing the number of copies of the minor allele in individual $n$ and SNP $q$. We assume that a weighted linear combination of the genotype data $X$ approximates the trait matrix $Y$ with Gaussian noise $\epsilon \in \Re^{N \times P}$:
\begin{eqnarray}
Y & = & X  Z  A + \epsilon \\
\epsilon_{n,\cdot} &\sim & N(0, \sigma_n^2 I)
\end{eqnarray}

As in a general reduced rank regression framework, we decompose the regression coefficients into a product of two matrices. Binary matrix $Z \in \{0,1\}^{Q\times K}$ selects sparse sets of SNPs for each of $K$ factors, and continuous matrix $A \in \Re^{K\times P}$ identifies subsets of traits that are associated with the SNPs included in one of $K$ factors. Our model diverges from the canonical RRR model in the prior distributions on the $A$ and $Z$ matrices. In particular, matrix $Z$ is a draw from the Indian Buffet Process (IBP) \cite{griffiths2005infinite,thibaux2007hierarchical}:
\begin{eqnarray}
z_{q,k} &\sim & Bernoulli(\pi_k) \\
\pi_k &\sim & Beta\left(\frac{\alpha}{K},1\right)\\
A_{k\cdot} &\sim & \N(0, \delta_{k\cdot} I ).
\end{eqnarray}
$K$ is inferred from the data, and is the number of nonzero columns in $Z$. \\

\begin{figure}
\begin{center}  
\includegraphics[scale=0.2]{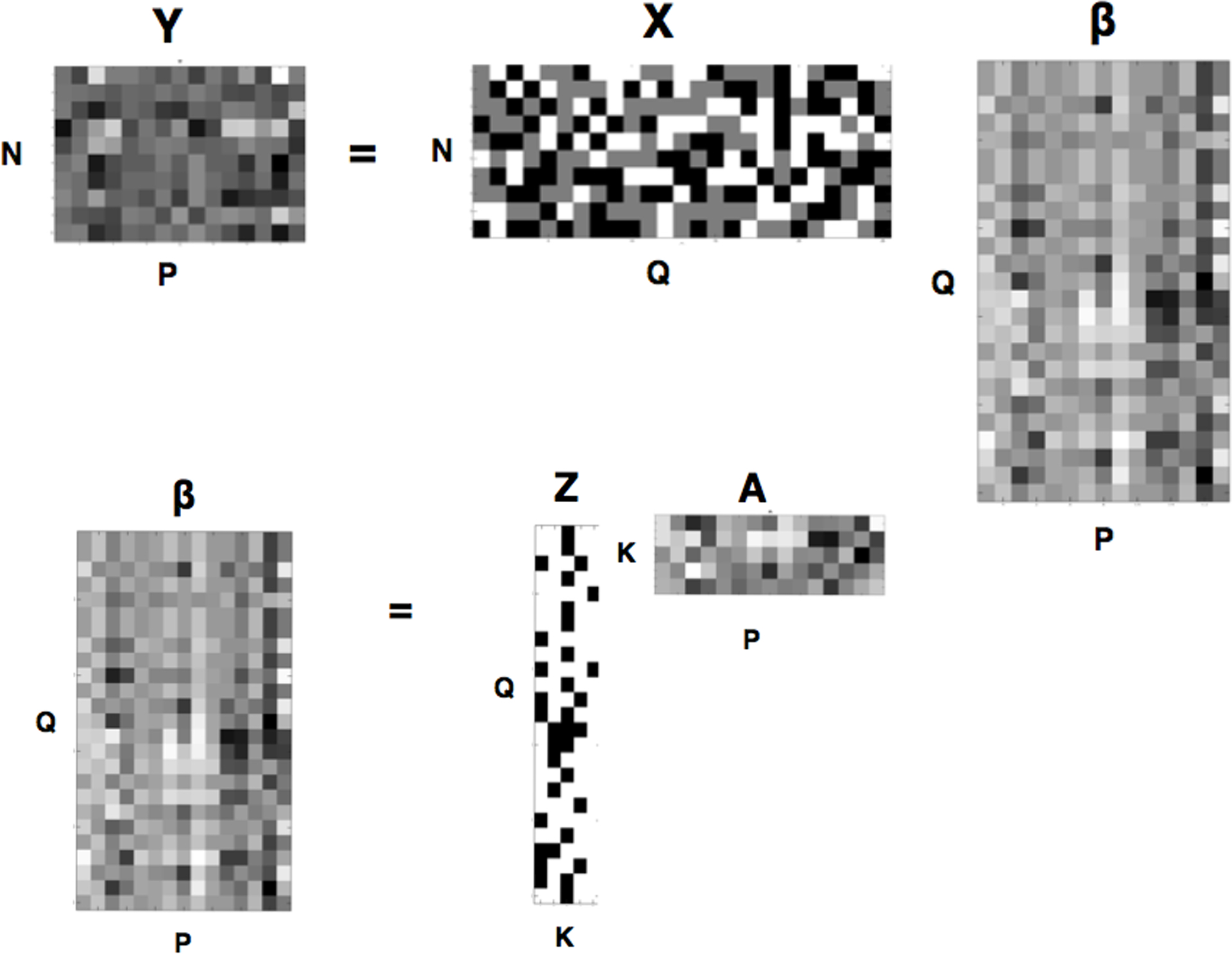}  
\caption{A visual representation of the BERRRI model. $Y$ is a matrix of $P$ correlated quantitative traits, $X$ is a genotype matrix $x_{n,q} \in {0,1,2}$, $\beta$ is the matrix of effects, decomposed into $Z$, the binary matrix, and $A$, the matrix of continuous effect sizes.
\label{fig:model}}
\end{center}  
\end{figure}

Moreover, we impose sparsity on the effect sizes, $A$, by placing a conjugate inverse-gamma prior on each parameter $\delta$, known as an automatic relevance determination (ARD) prior~\cite{neal_bayesian_1995, tipping_sparse_2001}. This has the effect of placing an $\ell_1$-type penalty on the effect sizes (Figure~\ref{fig:model}). 

The IBP is a Bayesian nonparametric prior on binary matrices with support on an infinite number of columns~\cite{griffiths2005infinite, thibaux2007hierarchical}. In the context of RRR, the IBP allows us to infer i) which SNPs are predictive of the observed data and ii) how many latent components $K$ are recovered (where each latent component is a separate multi-SNP, multi-trait association). The IBP prior has the desirable behavior that, in the posterior, we recover a finite number of components conditioned on a finite data set, and additional components may be added as more samples are collected to improve statistical power. 
Because $Z$ is binary, we may interpret the elements in a model selection context: the posterior probability of $z_{q,k}$ represents the probability that SNP $q$ is associated with the traits included in the $k$th component.

IBP estimation is a computationally difficulty problem, with an enormous number---$2^{q\times K}$---of binary configurations. Sampling methods for posterior estimation in the IBP are computationally intractable and sensitive to initialization~\cite{doshi2008variational}. These problems are magnified in high-dimensional genomic data. A variational Bayesian approximate method was recently developed for the IBP \cite{blei2006variational,doshi2008variational}. This variational approach exploits the stick breaking representation of the IBP~\cite{teh2007stick} for fast and robust posterior estimation. We base posterior inference in the BERRRI model on this variational approach.

\subsection*{Variational Bayes and the IBP}

Variational Bayes is an alternative to Monte Carlo sampling methods that provides an analytic solution to an approximation of the posterior distribution~\cite{attias1999inferring}. In contrast, sampling techniques provide an approximation to the exact posterior distribution. In many cases, variational Bayes methods provide solutions with greater computational efficiency than sampling methods, with minimal sacrifice in accuracy~\cite{teh2006collapsed}.

In the BERRRI model, the set of parameters to estimate are $W = {A, Z, \pi, \delta}$, and the fixed hyperparameters are $\theta = {\alpha, \sigma_n^2}$.
We write the log probability of $W$:
\begin{equation}
\log p(W | Y, X, \theta) = \log p(W, Y | X, \theta) - \log p(Y | X,\theta),
\end{equation}
where the log marginal probability:
\begin{equation}
\log p(Y | X,\theta) = \log \int p( Y, W | X, \theta) dW \label{eq:logmarginal}
\end{equation}
is intractable to compute.

Mean field variational Bayes methods approximate this log marginal probability with a variational distribution $q(W)$ \cite{beal2003variational,wainwright2008graphical}. 
We follow prior work~\cite{doshi2008variational} and use a finite IBP variational approach to these updates. 
Variational inference allows us to approximate model parameters $\lambda, \eta, \phi^m $ by minimizing the Kullback-Leibler divergence between an approximate variational distribution and a finite approximation to the IBP~\cite{doshi2008variational}. In our case, we use the following factorized variational distribution:
\begin{eqnarray}
q(W) = q_{\lambda_k}(\pi_k)q_{\eta_{qk}}(Z_{qk})q_{\phi_{kp},\varphi_{kp}}(A_{kp})q_{\kappa}(\delta)
\end{eqnarray}
where
\begin{eqnarray}
q_{\lambda_k}(\pi_k) &=& Beta(\pi_k;\lambda_{k,1},\lambda_{k,2}), \\
q_{\eta_{q,k}}(Z_{q,k}) &=& Bernoulli(Z_{q,k};\eta_{qk}) \\
q_{\phi_{k,p},\varphi_{k,p}}(A_{k,p}) &=& {\mathcal N}(A_{k,p};\phi_{k,p},\varphi_{k,p}) \\
q_{\kappa_{k,p}}(\delta_{k,p}) &=& Gamma^{-1}(\delta_{k,p};\kappa_{k,p,1},\kappa_{k,p,2})
\label{eq:factorized}
\end{eqnarray}

We then optimize the variational parameters to minimize the Kullback-Leibler divergence $D(q \mid \mid p)$ to the finite approximate IBP model distribution $p$. This reduces to an optimization problem of maximizing the lower bound on $p(Y | X,\theta)$. While this optimization can be difficult, when the distributions are in the exponential family there exists a closed form solution to the minimizing solution with respect to the variational parameters~ \cite{beal2003variational,wainwright2008graphical}. Minimizing this KL divergence is equivalent to maximizing the lower bound on the log likelihood (i.e., the evidence lower bound). Putting together Equations (\ref{eq:factorized}) and (\ref{eq:logmarginal}), we can optimize each of the variational parameters in closed form and conditionally on the others. In the fully factorized form, we can perform these updates sequentially; this produces a coordinate ascent optimization.

The optimal variational parameters $\varsigma_i$ that correspond to $W_i$ are the solution to:
\begin{eqnarray}
\log q_{\varsigma_i}(W_i) = E_{W_{-i}} [\log p(W,Y|X,\theta)] + C,
\end{eqnarray}
where the expectation is taken over all $W$ except $W_i$ according to the variational distribution. 

The updates for the variational parameters, $\lambda_{k,1}$ and $\lambda_{k,2}$ of the stick breaking weights $\pi$ are:
\begin{eqnarray}
\lambda_{k,1} &=& \frac{\alpha}{K} + \sum_{q=1}^Q \eta_{q,k} \label{eq:lam1}\\
\lambda_{k,2} &=& 1 + \sum_{q=1}^Q (1-\eta_{q,k}). \label{eq:lam2}
\end{eqnarray}

The updates for the variational parameters of the binary SNP-factor membership variables $Z$ are:
\begin{eqnarray}
\eta_{qk} &=& \frac{1}{1 + e^{-\zeta}} \label{eq:z} \\
\zeta &=& (\psi(\lambda_{k1}) - \psi(\lambda_{k2})) \\
&-& \frac{1}{2\sigma_n^2} \sum_{n=1}^N
(X_{nq} (tr(\varphi_{k}) + \phi_{k.} \phi_{k.}^T)X_{nq}^T) \notag \\
&-& \frac{1}{\sigma_n^2} \sum_{n=1}^N (X_{n,q} (tr(\varphi_{k}) + \phi_{k.} \phi_{k.}^T) \eta_{q^\prime k}^T X_{nq^\prime}^T \notag \\
&+& \frac{1}{\sigma_n^2} \sum_{n=1}^N X_{n,q} \phi_{k.} (Y_{n.}^T - \phi_{k^\prime.}^T \eta_{.k^\prime}^T X_{n.}^T) \notag
\end{eqnarray}

The updates for the variational parameters of effect sizes $A$ are:
\begin{eqnarray}
\varphi_{k} &=& (\frac{\kappa_{k \cdot 1}}{\kappa_{k \cdot 2}} + \frac{1}{\sigma_n^2} \sum_{n=1}^N \sum_{q=1}^Q X_{n.}\eta_{.k} X_{n.}^T)^{-1} \label{eq:avarva}\\
\phi_{k.} &=& \sum_{n=1}^N X_{n.}\eta_{.k} [(Y_{n.})^T - \sum_{k^\prime < k} (\phi^m_{k^\prime .})^T \eta_{.k^\prime}^T X_{n.}^T) \label{eq:ameanva}\\
&*& (\frac{\kappa_{k \cdot 1}}{\kappa_{k \cdot 2}} + \frac{1}{\sigma_n^2} \sum_{n=1}^N \sum_{q=1}^Q X_{n.}\eta_{.k} X_{n.}^T)^{-1} \notag
\end{eqnarray}

Finally, the updates for $\sigma_{A}^2$ are:
\begin{eqnarray}
\kappa_{k \cdot 1} &=& c + \frac{1}{2} \label{eq:va1}\\
\kappa_{k \cdot 2} &=& d + \sum_{k=1}^K \frac{\varphi_{kp} + \phi_{kp}^2}{2}. \label{eq:va2}
\end{eqnarray}
Full derivations of the variational updates are available in the Supplementary material. 

To assess convergence, we used the following approach, inspired by the Geweke test statistic~\cite{geweke1991evaluating}. Let $m_{i,1}$ and $s_{i,1}$ and be the mean and the standard deviation, respectively, of the first 10\% of the trace of $\varsigma_i$ (i.e., estimates of $\varsigma_i$ over $n_1$ iterations). Let $m_{i,2}$ and $s_{i,2}$ be the mean and standard deviation of the last 50\% of the trace of $\varsigma_i$ (i.e., estimates of $\varsigma_i$ over $n_2$ iterations). Both percentages exclude 100 burn-in iterations. We assessed convergence every 100 iterations by checking for unequal variance for all $\varsigma_i$, where the t-statistic $t_i$ equals:

\begin{eqnarray}
t_i &=& \frac{m_{i1}-m_{i2}}{\sqrt{\frac{s_{i1}}{n_1} + \frac{s_{i2}}{n_2}}} \label{eq:convergei}
\end{eqnarray}

P-values $p_i$ are computed for each parameter type $t_i$. If $p_i \leq p_{thresh}$, where we set $p_{thresh} = 0.05$, the segments of that trace are significantly different and the model has not yet converged. We find that, in general, this variational approach converged within $500$ iterations. 

The algorithm for fitting the BERRRI model is included here (Algorithm~\ref{BERRRIsalgorithm}).
\begin{algorithm}
\caption{BERRRI algorithm}
\label{BERRRIsalgorithm}
\begin{algorithmic}[1]
\Procedure{Fit BERRRI}{}
\While{$i$ \textless $maxiter$ and $p_i \leq p_{thresh}$ for at least one parameter} 
\For{$k$ = 1...$K$}
\State Update $\lambda_{k,1}$ with Equation \ref{eq:lam1}
\State Update $\lambda_{k,2}$ with Equation \ref{eq:lam2}
\EndFor
\For{$k$ = 1...$K$}
\For{$q$ = 1...$Q$}
\State Update $\eta_{q,k}$ with Equation \ref{eq:z}
\EndFor
\EndFor
\For{$k$ = 1...$K$}
\State Update $\varphi_k$ with Equation \ref{eq:avarva}
\State Update $\phi_k$ with Equation \ref{eq:ameanva}
\EndFor
\For{$k$ = 1...$K$}
\For{$p$ = 1...$P$}
\State Update $\kappa_{k,p,1}$ with Equation \ref{eq:va1}
\State Update $\kappa_{k,p,2}$ with Equation \ref{eq:va2}
\EndFor
\EndFor
\EndWhile
\EndProcedure
\end{algorithmic}
\end{algorithm}

\subsection*{Testing for eQTLs}

To assess whether a SNP is an eQTL for a specific gene, we computed variational maximum a posteriori (VMAP) estimates for each SNP-gene pair. According to our model, a SNP is an eQTL for a gene if the SNP is in any factor K \emph{and} factor K has a non-zero weight for the observed gene expression levels. 
We can interpret an estimate of variational parameter $\eta_{q,k}$ as the posterior probability that SNP $q$ is included in factor $k$. We therefore compute the matrix of $V_{MAP}$ estimates for SNP-gene pairs as follows:
\begin{eqnarray}
V_{MAP} = E_q[Z]E_q[A] = \eta \phi \label{eq:bf}
\end{eqnarray}
where $\E_q[A]$ and $\E_q[Z]$ are the expected value of parameters $A$ and $Z$ under the variational distributions. 

To determine a level at which a $V_{MAP}$ for a SNP-gene association should be considered significant, we performed permutation testing to identify a $V_{MAP}$ threshold with a false discovery rate (FDR) of $\leq 10\%$.

\section*{Methods}

\subsection*{Simulated Data}

Simulated data were generated for $N = \{25,100,500\}$ individuals, $P = \{25\}$ genes, $Q = \{25,100,1000\}$ SNPs, and $K = \{5,25\}$ latent factors. These data were generated using actual genotype data from HapMap Phase 3, chromosome 20. For each simulated data set, a random subset of $N$ individuals and $Q$ SNPs were selected from the genotype data. An inclusion matrix $Z$---indicating which SNPs are included in each factor---was simulated by, for each $k \in \{1,\dots,K\}$, selecting a SNP at random to include in the factor, and then including additional SNPs based on the correlation of the genotypes across all individuals with that of the index SNP.
A weight matrix $A$---encoding the effect size that a group of SNPs has on each of a subset of traits---was then simulated from a normal distribution with mean zero and standard deviation $0.5$.

The gene expression matrix $Y$ was then computed from the simulated parameters, and independent Gaussian random noise was added to each element with mean zero and standard deviation $1$.

\begin{eqnarray}
\epsilon_{i,j} \sim {\mathcal N}(0,1) \\
Y = XAZ + \epsilon
\end{eqnarray}
The code to generate simulated data is included as Supplementary material.

\subsection*{Methods for comparison}

We compared the results of BERRRI on these simulations with two other RRR methods: SRRR~\cite{chen2012sparse} and CRAM \cite{foygel2012nonparametric}. 
SRRR is a penalized least-squares RRR method for simultaneous dimension reduction and variable selection. Dimension reduction is accomplished by explicitly representing the $Q$ by $P$ matrix of regression coefficients as two rank $K$ matrices. SRRR accomplishes variable selection with a group lasso penalty \cite{yuan2006model}, determining optimal parameters using the subgradient method \cite{friedman2007pathwise}.

CRAM uses an additive model to estimate each dimension $P$ of the response, constraining the complexity of the model with the nuclear norm and encouraging the set of functions to be low rank~\cite{foygel2012nonparametric}. The authors derive backfitting algorithms to estimate optimal parameters. Compared to RRR methods that use an orthogonal projection onto a lower rank space, the CRAM method is well defined for high-dimensional data.

SRRR requires two hyperparameters to be specified: $\lambda$, the penalty parameter, and $K$, the rank of the RRR. CRAM requires the specification of a single penalty parameter, $\lambda$. For both SRRR and CRAM, we estimated hyperparameters via five-fold cross-validation. Each method was subsequently run on the simulated data sets ten times using the optimal hyperparameters. 

The computational complexity for both CRAM and SRRR is $O(NKQ^2)$ with some smaller linear terms, where $Q$ is the dimension of the explanatory matrix, $N$ the number of observations, and $K$ the rank of the RRR. The computational complexity of BERRRI is $O(NPK^2Q^2)$ with some smaller linear terms, making clear the cost of the additional representation in terms of computational time.

\subsection*{Evaluation}

For each method and each simulation, we measured the overall wall clock run time, residual sum of squares (RSS), and the precision and recall for the true associations recovered from the simulated data. The 95\% confidence interval for each metric was computed. 

\subsection*{HapMap Phase 3 eQTL study}

We applied BERRRI to a subset of data from the HapMap Phase 3 study. These data include genotype data from $608$ individuals from $11$ worldwide populations~\cite{gibbs2003international}. Gene expression microarray data were collected on the lymphoblastoid cell lines (LCLs) from these $608$ individuals~\cite{Stranger2012}. We processed these microarray data to remove low expressed genes and technical and biological effects that may confound association~\cite{Engelhardt2014}. We projected each of the gene expression levels to the quantiles of a standard normal to control for non-Gaussianity and outliers; the full data processing pipeline is described elsewhere~\cite{Brown2013}.

We considered all expressed genes located on chromosome 21, and selected at random 1\% of all non-redundant SNPs on the chromosome. We applied BERRRI to the gene expression matrix $Y \in \Re^{608 x 175}$ and genotype matrix $X \in \Re^{608 x 171}$. We ran the variational inference algorithm until convergence,
and used the point estimate of the parameters as our result. Finally, the variational map estimates for associations of each SNP with each gene were computed (Equation \ref{eq:bf}). 

For comparison purposes, we also computed univariate Bayes factors for each SNP $Q$ and gene $P$ pair using BIMBAM ($BF_{U}$)~\cite{servin2007imputation}. Bayes factor 10\% FDR thresholds for both approaches were identified via permutation testing.

\subsection*{Global FDR Thresholds}
We determined the $BF_{U}$ and VMAP thresholds for FDR $\leq 10\%$ using permuted data. More specifically, We fit the BERRRI model and computed the VMAP using (Equation \ref{eq:bf}) above for the true data, and then fit the BERRRI model and computed the VMAP again with permuted values of the response (i.e., shuffling sample labels of the gene expression matrix). We used the permuted VMAP estimates as a null VMAP distribution, assuming no associations. The same permuted data is used to compute a null distribution of $BF_{U}$.

Using these null distributions, we are able to compute a false discovery rate (FDR) for various thresholds of $BF_{U}$ and VMAP. The FDR for a given threshold for each method is computed as the number of BFs or VMAPs exceeding the threshold in the permuted data (i.e., the number of false positives), divided by the number of BFs or VMAPs exceeding the threshold in the true data (i.e., the number of true positives plus false positives), scaled by the relative number of tests in the permuted and true data. For each method, we identified the threshold having less than or equal to 10\% FDR to determine a cutoff for significant associations.

\section*{Discussion}

\subsection*{Simulation results}

\begin{figure}
\begin{center} 
\includegraphics[height=3.5in,angle=0]{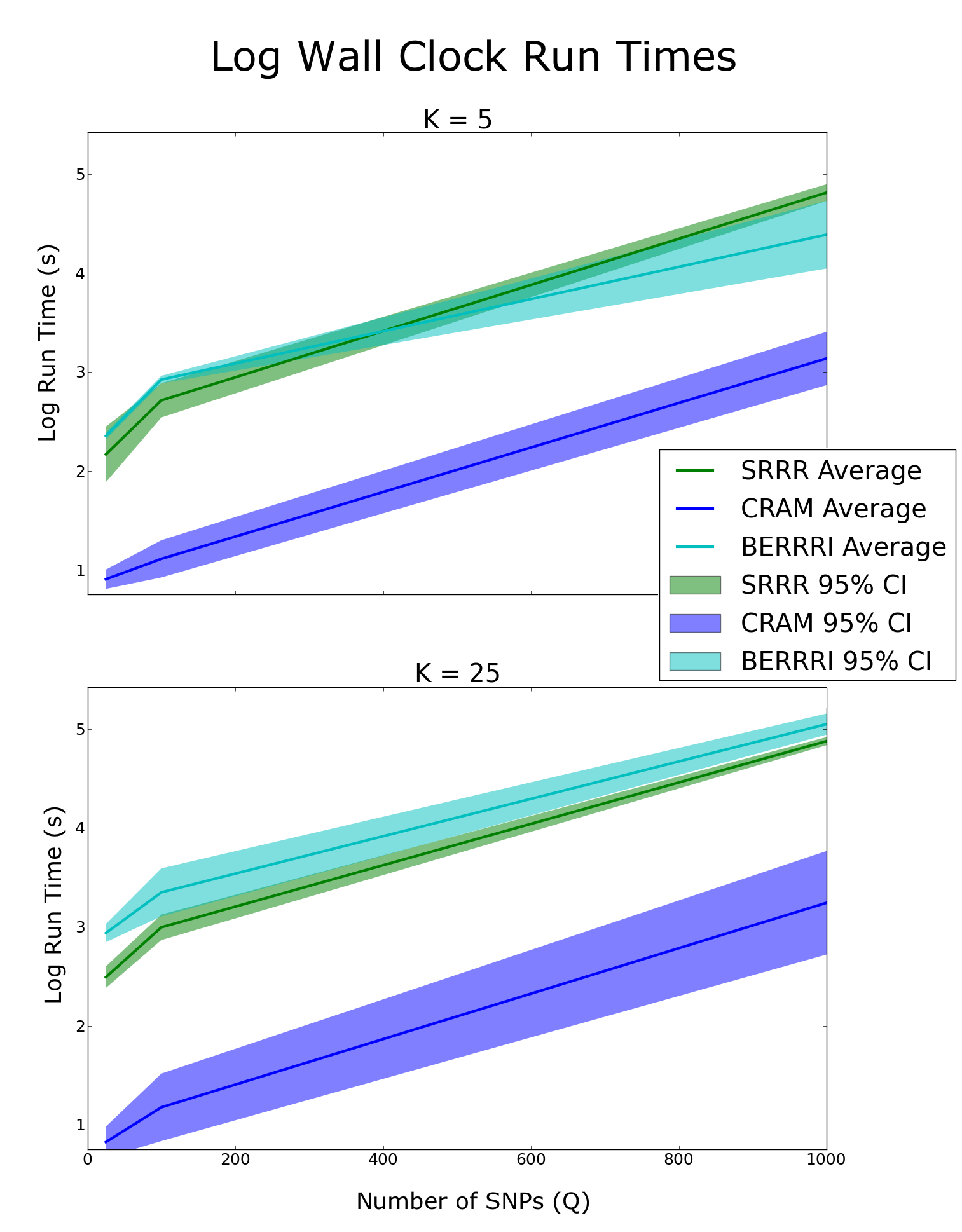}  
\caption{\small \sl {\bf Overall log wall clock run time in seconds for CRAM, SRRR, and BERRRI.} x-axis: number of SNPs in the test; y-axis: log wall clock time. The dark lines show the average across ten runs; the ribbons show plus or minus three standard deviations from the mean run time. \label{fig:simRunTimes}}  
\end{center}  
\end{figure}

To illustrate the effectiveness of the BERRRI approach, we simulated multiple correlated quantitative traits and genotype data in order to recover the associations among multiple SNPs and multiple traits. We compared our model to SRRR \cite{chen2012sparse} and CRAM \cite{foygel2012nonparametric} (Figure ~\ref{fig:simRunTimes} and Table ~\ref{Tab:02}).

\begin{table}[!t]
\centering
\caption{Residual sum of squares (RSS) for the predicted trait values versus true response values in matrix $Y$ for CRAM, SRRR, and BERRRI. Average and standard deviation are computed from ten runs. $K$ represents the number of simulated factors.\label{Tab:02}}
\begin{tabular}{lllll}\hline
Method & SNPs ($Q$) & True $K$ & Avg. RSS & Std. Dev. \\
\hline
SRRR & 25 & 5 & 1777.72 & 459.82 \\
CRAM & 25 & 5 & {\bf 832.14} & $1.9\times 10^{-3}$ \\
BERRRI & 25 & 5 & 1921.60 & 103.54 \\
\hline
SRRR & 25 & 25 & 4,298.29 & 1,389.94 \\
CRAM & 25 & 25 & 4,160.82 & $8.9\times 10^{-3}$ \\
BERRRI & 25 & 25 & {\bf 3,343.82} & 227.64 \\
\hline
SRRR & 100 & 5 & 4,101.31 & 1,246.95 \\
CRAM & 100 & 5 & 7,175.40 & $2.6\times 10^{-7}$ \\
BERRRI & 100 & 5 & {\bf 1,095.64} & 100.02 \\
\hline
SRRR & 100 & 25 & 15,611.5 & 5,198.9 \\
CRAM & 100 & 25 & 41,331.5 & $1.2\times 10^{-6}$ \\
BERRRI & 100 & 25 & {\bf 2,707.07} & 505.0 \\
\hline
SRRR & 1000 & 5 & {\bf 97,798} & 32,477 \\
CRAM & 1000 & 5 & 534,081 & $1.1\times 10^{-5}$ \\
BERRRI & 1000 & 5 & 547,049 & 271,977 \\
\hline
SRRR & 1000 & 25 & 238,136 & 79,350 \\
CRAM & 1000 & 25 & 4,528,967 & $4.9\times 10^{-5}$ \\
BERRRI & 1000 & 25 & {\bf 157,692} & 193,555 \\\hline
\end{tabular}
\end{table}

\begin{figure}
\begin{center}  
\includegraphics[height=2.75in]{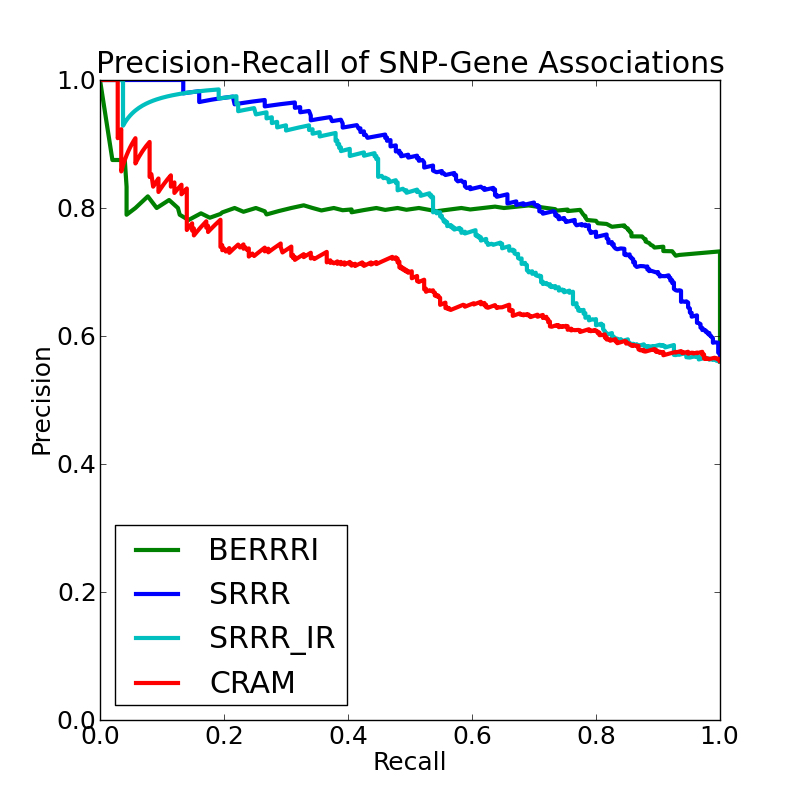}  
\caption{\small \sl Estimated precision and recall of CRAM, SRRR, SRRR with an incorrectly specified rank (SRRR\_IR), and BERRRI. Precision and recall quantify the proportion of discovered associations that are truly significant and the proportions of truly significant associations that are discovered, respectively, quantified using simulated data with ground truth and the variational MAP estimates of effect size. 
\label{fig:prc}}  
\end{center}  
\end{figure}

BERRRI performs well relative to existing RRR methods CRAM and SRRR in the prediction of held out response variables (Figure~\ref{fig:prc}; Table~\ref{Tab:02}). SRRR outperforms BERRRI and CRAM for the RSS metric in one scenario when the simulated rank is low. This may due to SRRR requiring a prespecified rank, which we set to the true rank of the simulated data, giving SRRR an advantage.

The overall run time for BERRRI and SRRR are similar, with CRAM being the fastest. All scale approximately exponentially with the number of simulated predictors $Q$ (Figure~\ref{fig:simRunTimes}), which makes association testing across all SNPs infeasible without further approximations.    

In order to quantify the proportion of true associations recovered using RRR models, we computed the precision-recall for each method based on the true associations in the simulated data (Figure~\ref{fig:prc}). At high levels of recall ($0.75$ and greater), BERRRI shows dominant precision around $0.75$; the other methods drop off in their precision much more sharply as the recall increases from $0.75$ to $1.0$. SRRR initially shows higher precision than the other methods, with a steady decline and falling below BERRRI near a $0.75$ recall rate. For this analysis, we also ran SRRR with an incorrectly specified rank (SRRR\_IR). As expected, this shows worse performance with a sharper decline in precision compared to SRRR. CRAM shows lower precision than the other methods for most recall rate levels. BERRRI recapitulates more of the underlying structure than SRRR and CRAM in the simulated data, while maintaining specificity and avoiding the detection of spurious associations.
\subsection*{Results on HapMap Phase 3 eQTL study data}

To illustrate the effectiveness of the BERRRI approach in multi-SNP, multi-trait association mapping, we applied BERRRI to a subset of data from the HapMap Phase 3 Study~\cite{gibbs2003international}. We selected chromosome 21, choosing at random 1 \% of all non-redundant SNPs, and all expressed genes on the chromosome. This resulted in a gene expression matrix $Y \in \Re^{608 x 175}$ and genotype matrix $X \in \Re^{608 x 171}$. We ran BERRRI on these data, and computed the BERRRI VMAP values to quantify associations of each SNP with each gene. We computed FDR using permutations. 

\begin{figure}
\begin{center}  
\includegraphics[height=2.75in,angle=0]{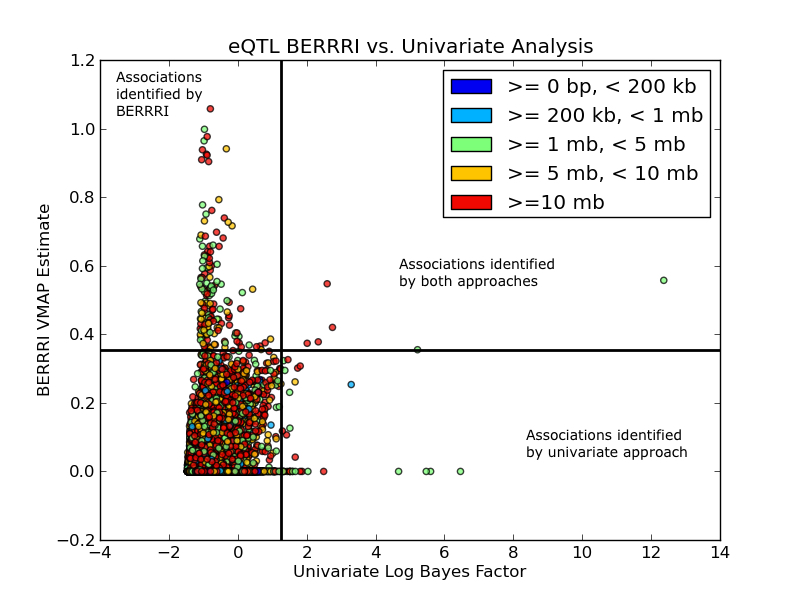} 
\caption{\small \sl BERRRI VMAP estimates for associations of each SNP with each gene from our model, versus Bayes factors from a classical eQTL univariate analysis \cite{servin2007imputation} on Hapmap Phase 3 eQTL study data. The vertical line indicates the 10\% FDR threshold for BERRRI, and the horizontal line indicates the 10\% FDR for the classical univariate analyses. Points in the upper left quadrant are SNP-gene associations significant only in the BERRRI approach. Points in the upper right quadrant are associations significant in both approaches. Associations in the lower right quadrant are only significant in the classical univariate approach. \label{fig:qq}}  
\end{center}  
\end{figure}

To determine whether or not BERRRI identifies true associations, we compared the BERRRI $VMAP$ values for the SNP-gene associations with the univariate Bayes factors ($BF_{U}$) for each SNP-gene pair~\cite{servin2007imputation} (Figure~\ref{fig:qq}).
We observed more eQTL associations using BERRRI (n=157 at 10\% FDR) than using the univariate approach (n=38 at 10\% FDR). By analyzing the gene expression and SNP data jointly in BERRRI and by reducing the dimensionality of the genotype data via latent factors, we were able to obtain additional statistical power in association mapping. This additional power will lead to improvements in discovering eQTLs with smaller effect sizes relative to the univariate approach.

Importantly, BERRRI fails to discover some associations found to be significant with the univariate analysis (Figure~\ref{fig:qq}). It could be that these missing associations are either false negatives in the univariate analysis or missing (singleton) true associations. Statistically, this occurs because a SNP is excluded from a reduced rank association, which may happen when the association is explained by another highly correlated SNP. 

We looked at the enrichment of proximal and distal associations identified by BERRRI and the univariate approach (Figure~\ref{fig:qq}). To assess whether BERRRI identifies more distal associations than the classical approach, we  performed a simple logistic regression, testing the hypothesis that distance between a SNP and a gene is significantly different for the associations identified by BERRRI only (upper left quadrant of Figure~\ref{fig:qq}), and the associations identified by the univariate approach only (lower right quadrant). We found that there was a significant enrichment (p = 0.0239) for more distal associations in the eQTLs identified by BERRRI.

\begin{figure}
\begin{center}  
\includegraphics[height=4.4in,angle=0]{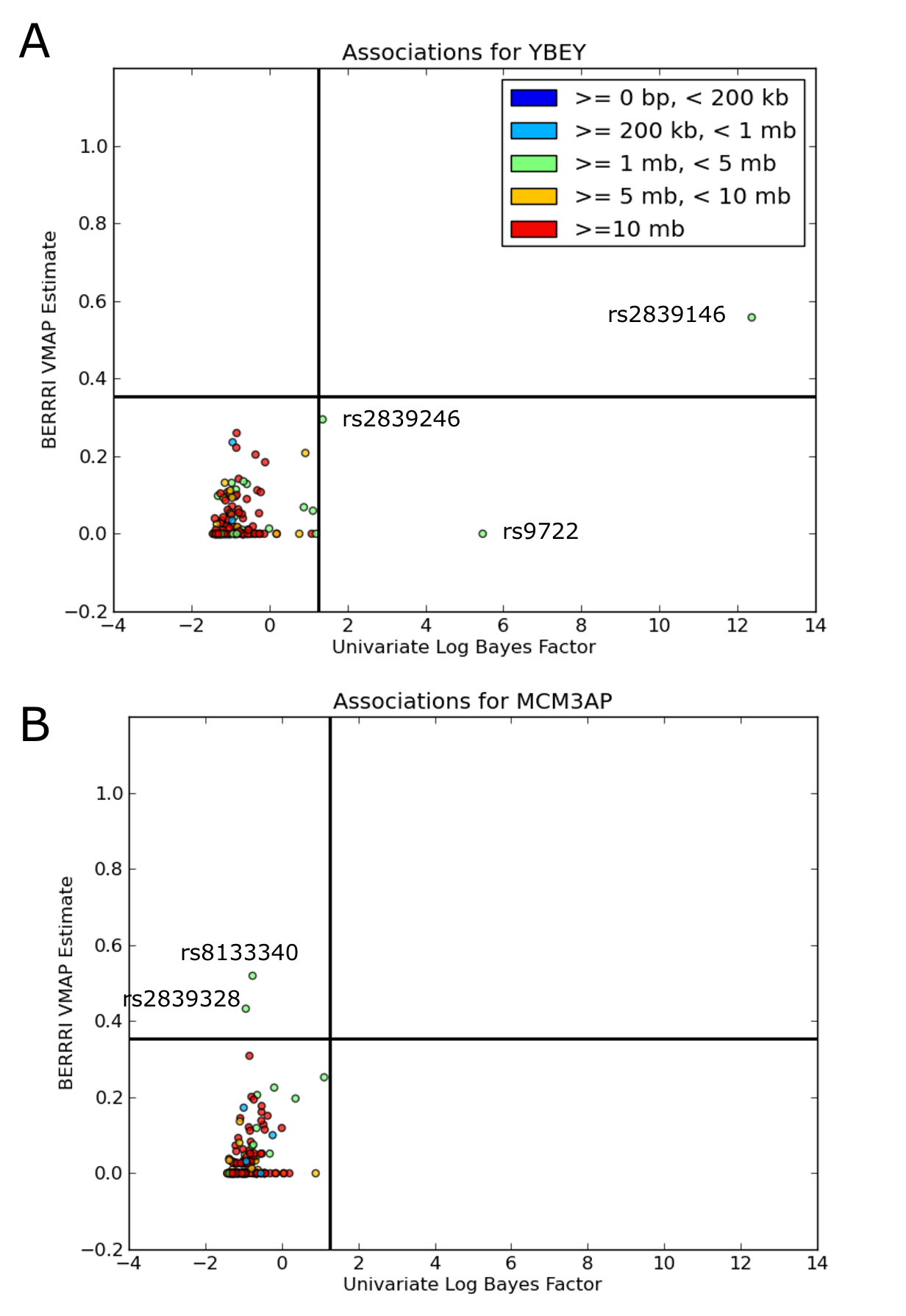}  
\caption{\small \sl The VMAP values from BERRRI and univariate approaches are shown on the vertical axes, with distance from the gene shown on the horizontal axis. The horizontal dashed lines correspond to the 10\% FDR for each method. Panel A: Associations identified for \textit{YBEY}. Panel B: Associations identified for \textit{MCM3AP}. \label{fig:gene1}}  
\end{center}  
\end{figure}

The most significant eQTL from the univariate approach is SNP rs2839146 for gene \textit{YBEY}, which is located 75.7 kb upstream of the gene transcription start site (TSS); the BERRRI VMAP value for this association is $0.56$ and the univariate log BF is $12.37$ (Figure~\ref{fig:gene1}A). This eQTL is significant in both approaches, and was identified as statistically significant across twelve tissues in the Genotype-Tissue Expression (GTEx) v6 data~\cite{ardlie2015genotype}. In the GTEx data, this SNP is also an eQTL for gene \textit{MCM3AP}. It is located 24.5 kb away from the gene TSS, and was identified as statistically significant across thirteen tissues in GTEx. The BERRRI VMAP value for this association is 0.25 and the univariate log BF is 1.11 (Figure~\ref{fig:gene1}bB), both trending toward significance. BERRRI identified two other potential eQTLs for this gene: rs2839328 (329.4 kb away), and rs8133340 (331 kb away) with VMAP estimates of $0.52$ and $0.43$, respectively. Neither of these SNPs were listed as eQTLs for \textit{MCM3AP} in GTEx, nor were they in LD with the GTEx listed eQTLs ($r^2 > 0.5$). 

The second most significant eQTL from the univariate approach is SNP rs3992 for gene \textit{PIGP}, which is located 74.5 kb from the gene TSS; the BERRRI VMAP value for this association is 0.0 and the univariate log BF is 6.47. This eQTL was identified as a statistically significant eQTL across twenty tissues in the GTEx data.

Another eQTL identified by both approaches is rs9974970 for gene \textit{HLCS}, which is located 217.7 kb from the gene TSS; the BERRRI VMAP value for this association is 0.35 and the univariate log BF is 5.22. This eQTL was identified as statistically significant across sixteen tissues in the GTEx data~\cite{ardlie2015genotype}.

BERRRI also identified eQTLs not discovered by the univariate approach. For example, SNP rs2839328 is identified as an eQTL for genes \textit{FTCD} (428.3 kb), and \textit{PRMT2} (40.5 kb), which were both significant eQTLs in the GTEx data. The BERRRI VMAP estimates are 0.53, and 0.61, respectively. This SNP is also identified by BERRRI as an eQTL for many other genes, including \textit{ADAMTS1}, \textit{LINC00160}, \textit{RBM11}, \textit{TRAPPC10}, and \textit{USP16}, all of which have a BERRRI VMAP estimate higher than 0.90, and all of which were outside the GTEx cis-eQTL testing window of 1 Mb away from a gene TSS. The univariate approach was underpowered to detect these associations.

In many of these SNP-gene results, the associations discovered by BERRRI results in greater locus heterogeneity, meaning that the associated SNPs are spread out across a wider range of cis-loci; this behavior has been observed in other association models that similarly separate estimation of SNP association and effect size, as we do here~\cite{efron2008microarrays,Engelhardt2014}. For \textit{SH3BGR}, BERRRI identified an association found by the univariate approach with more significance, and identified one association that the univariate approach was underpowered to find. Similarly, with \textit{ADAMTS5}, BERRRI identified an eQTL that went undetected by the univariate approach. Both of these SNP-gene pairs were outside the 1 Mb cis-eQTL testing window of GTEx.

For \textit{ETS2} and \textit{PCNT}, a few nominally significant associations detected by the univariate approach went undiscovered by BERRRI. We note that the undiscovered associations may either be a true association that is missed by BERRRI or a false positive univariate association that BERRRI avoids by sharing strength across genes and SNPs.

Imposing additional sparsity on the effect size matrix could further improve interpretation and association testing. For instance, a two-groups model such as a spike-and-slab Bayesian prior on $A$ would result in an explicit posterior probability of association for eQTLs. This improvement may alleviate the challenge of detecting small effect eQTLs~\cite{efron2008microarrays}.

\section*{Conclusion}

In this work, we developed a Bayesian nonparametric reduced rank regression method that effectively estimates a high-dimensional response matrix, inferring low-rank dimensionality from the data, and produces an interpretable model without any significant increase in computational time over state-of-the-art methods. While the results here illustrate an application to eQTL analyses, BERRRI may be applied to any multi-dimensional association analysis. We showed in simulations that our model accurately recovers latent structure in the response matrix and identifies associated predictors, improving over existing RRR methods lacking the interpretability and nonparametric behavior of the BERRRI approach. In the context of eQTL studies, BERRRI shows improved statistical power to identify genetic associations compared to classical univariate tests. 

\section*{Acknowledgement}
The authors would like to acknowledge Dr. Rina Foygel for providing CRAM software. This work was supported by the National Institutes of Health [R00 HG006265 and NIH R01 MH101822 to BEE]; and the Defense Advanced Research Projects Agency Predicting Health and Disease Award [N66001-09-C-2082 to GSG]. AMV was supported by the Gates Foundation [OPP1017554 to GSG].

\bibliographystyle{plain}
\bibliography{document.bib}

\end{document}